\documentclass[aps,prx,twocolumn,superscriptaddress,showpacs]{revtex4}

\usepackage{amsmath, amsfonts, amssymb, graphicx,epstopdf,float}

\newcommand{\ket}[1]{| #1 \rangle}

\newcommand{\braketop}[3]{\langle #1 | #2 | #3 \rangle}
\newcommand{\ave}[1]{\left\langle #1 \right\rangle}
\newcommand{\dg}[0]{\dagger}

\newcommand{\vk}[0]{{\bf k}}

\newcommand{\vq}[0]{{\bf q}}
\newcommand{\vQ}[0]{{\bf Q}}

\newcommand{\vR}[0]{{\bf R}}

\renewcommand{\vr}[0]{{\bf r}}

\renewcommand{\epsilon}{\varepsilon}
\newcommand{\up}[0]{{\uparrow}}
\newcommand{\dn}[0]{{\downarrow}}

\begin{document}
\title{Probing Competing and Intertwined Orders with Resonant Inelastic x-ray Scattering in the Hole-Doped Cuprates}

\author{David Benjamin}
\affiliation{Physics Department, Harvard University, Cambridge, Massachusetts, USA
}

\author{Israel Klich}
\affiliation{Department of Physics, University of Virginia, Charlottesville, VA, USA}

\author{Eugene Demler}
\affiliation{Physics Department, Harvard University, Cambridge, Massachusetts, USA
}

\date{\today}

\begin{abstract}
We develop a formalism to study indirect resonant inelastic x-ray scattering (RIXS) in systems with itinerant electrons, accounting for the attraction between valence electrons and the positively-charged core hole exactly, and apply this formalism to the hole-doped cuprate superconductors.    We focus on the relationship between RIXS lineshapes and band structure, including broken symmetries.  We show that RIXS is capable of distinguishing between competing order parameters, establishing it as a useful probe of the pseudogap phase.
\end{abstract}


\maketitle

\section{Introduction}
The state of the underdoped cuprates above the superconducting transition temperature is an outstanding puzzle in the field of high-temperature superconductivity.  Many different types of order~\cite{Timusk1999,Kivelson1998,Sachdev2000,Chakravarty2001,Kivelson2003a,Norman2005,Valla2006a,Vojta2009b,Berg2009,Baeriswyl2009,Davis2013} have been hypothesized to coexist or compete with superconductivity and to explain pseudogap behavior in which the density of states is depleted around the Fermi energy.  These potential phases are difficult to detect because, with the exception of charge density wave (CDW) and spin density wave (SDW) in certain cuprates at specific dopings, they exhibit dynamic fluctuations and spatial inhomogeneity~\cite{Kivelson2003a, Sachdev2003}.  These fluctuations smear the signatures of potential order parameters and render them undetectable by quasistatic probes that effectively average over time scales longer than those of fluctuations.  Resonant inelastic x-ray scattering (RIXS) can overcome this obstacle because its time scale  is bounded by the finite lifetime (roughly $(250 \, {\rm meV})^{-1} \approx 1 \, {\rm fs}$) of an intermediate state core hole due to Auger decay and other processes.  Essentially, RIXS can take instantaneous snapshots of fluctuating short-range order. However, there exists little theory for RIXS in systems with itinerant electrons; in particular, it is not clear how RIXS can be used to probe competing orders, such as those that appear (and are hypothesized to exist) in the normal phase of the cuprate superconductors.  In this paper we show that indirect RIXS, in which one measures the effect of a positively-charged transient core hole on the valence band Fermi sea, is sensitive to changes in the particle-hole excitation spectrum induced by many of the most commonly considered order parameters proposed for the peudogap phase.   We show that it is possible to detect and distinguish different order parameters, for example charge density waves, antiferromagnetism, and the hypothesized $d$-density wave~\cite{Chakravarty2001}.  We derive a formula for indirect RIXS in systems with itinerant electrons, accounting exactly for the core hole potential acting on valence electrons.

\section{Theoretical Formalism} 
In indirect RIXS a photon is absorbed and causes a transition of a core electron to an excited band \textit{above} the valence band, leaving behind a positively-charged core hole.  In the cuprates, the most commonly-used transition is $1s \rightarrow 4p$.  The excited band is generally weakly-interacting; in cuprates this occurs because it is derived from a delocalized $4p$ orbital.  Thus the excited electron does very little in indirect RIXS except re-fill the core hole and emit a scattered photon.  The interesting action is the effect of the core hole on the electrons in the valence band, which in a conducting system is to generate particle-hole ``shake-up'' pairs.  When one measures the energy difference $\Delta \omega$ and momentum difference $\Delta \vq$ between the incident and scattered photons, one is measuring the dispersion of shake-up pairs.  The spectral weights of shake-up processes at different energy and momenta reveal the joint density of states of particles and holes, which in turn sheds light on quasiparticle dispersions modified by different types of order.

The intensity for incident x-rays of momentum $\vq_i$ and energy $\omega_i$ to be scattered into outgoing momentum $\vq_f=\vk_i+\Delta \vk$ and energy $\omega_f=\omega_i-\Delta \omega$ is calculated from the familiar Kramers-Heisenberg formula~\cite{Kotani2001}
\begin{align}
\label{KramersHeisenberg}
I \propto & \sum_f \left| \sum_n
\frac{     \braketop{f}{T}{n} \braketop{n}{T^\dg}{i}     }{     \omega_i+E_n-E_i+i \Gamma     }
\right|^2 \delta(E_f-E_i-\Delta \omega),
\end{align}
where $\ket{i(n,f)}$ are the initial (intermediate, final) states with energies $E_{i(n,f)}$, $1/\Gamma$ is the lifetime of the intermediate state core hole, and $T^\dg = \sum_m e^{\vq_i \cdot \vR_m}s_m p^\dg_m$, $T = \sum_m e^{-\vq_i \cdot \vR_m} s^\dg_m p_m$ are transition operators.  This notation reflects the $1s \rightarrow 4p$ indirect RIXS in the high-T$_c$ cuprates.   We assume momentum-independent dipole matrix elements and leave polarization dependence implicit~\cite{Abbamonte2012}.  The attractive potential due to the core hole is most easily accounted for by switching to the time domain~\cite{Nozieres1974, Benjamin2013a}, where we have
\begin{align}
\label{TimeDomain}
I \propto & \sum_{m,n}e^{i \Delta \vq \cdot (\vR_n-\vR_m)} \int_{-\infty}^\infty ds \int_0^{\infty} dt  \int_0^{\infty} d\tau \nonumber \\
& e^{i\omega_i(t-\tau)-i s \Delta \omega-\Gamma (t+\tau)} S_{mn},
\end{align}
where
\begin{equation}
\label{Smn}
S_{mn} =
\ave{e^{i H_0 \tau} p_n e^{-i H_n \tau} p^\dg_n e^{i H_0 s} p_m e^{i H_m t} p^\dg_m e^{-i H_0(t+s)}}.
\end{equation}
Eq.~(\ref{Smn}) is a Keldysh-like integral describing the history of absorption and emission events separated by time evolution operators.  Since the core hole is immobile and has a constant energy that can be absorbed into $\omega_i$ we are able to remove via $H \rightarrow H_{m(n)} \equiv H_0+V_{m(n)}$, where $H_0 = H_d + H_p$ is the Hamiltonian of valence $d$ electrons and the $p$ band and $V_m$ is the potential due to the core hole at site $m$ that acts on valence electrons.  As the $4p$ band is highly dispersive we assume that $4p$ electrons do not interact with the core hole or the valence band; this is the usual ``spectator'' approximation~\cite{Tsutsui1999}.  Therefore, the $d$ and $p$ bands are separable and the $4p$ contribution to Eq. (\ref{Smn}) reduces to a product of Green functions:
\begin{align}
&S_{mn} = G^{nn}_p(\tau) G^{mm}_p(-t) \times \nonumber \\
\label{IndirectS}
& \quad \ave{e^{i H_d \tau}  e^{-i H_{d,n} \tau}  e^{i H_d s}  e^{i H_{d,m} t}  e^{-i H_0(t+s)}},
\end{align}
where $G^{nn}_p(t) = \braketop{0}{p_n(t)p^\dg_n(0)}{0}$ is an easily-calculated single-particle quantity.  We note that by Eq.~(\ref{TimeDomain}) $S_{mn}$ is measurable as the Fourier transform of the intensity.  Given the form of $S_{mn}$ as a Keldysh-like correlator it is intuitively clear that it may reveal the real-space structure of $H$.  We will discuss this point further below.

Following a recent analysis of direct RIXS~\cite{Benjamin2014} we treat the valence band as a system of non-interacting quasiparticles, which is valid when the quasiparticle lifetime is long compared to the core hole lifetime $1/\Gamma$.  In the cuprates $\Gamma \ge 250$ meV, which exceeds quasiparticle widths even quite far from the the Fermi surface.  Many-body averages of products of exponentiated quadratic operators such as in Eq.~\ref{IndirectS}  have been discussed in numerous works~\cite{Scalapino1981, Scalettar1986, Klich2004, Abanin2005, Benjamin2013}.  The standard formula $\ave{e^Z} = \det \left[(1-\hat{N}) + e^z \hat{N} \right]$, where uppercase `$Z$' denotes a quadratic many-body operator and lowercase `$z$' denotes its matrix elements $Z = d^\dg_i z_{ij} d_j$ and $\hat{N}$ is the Fermi occupation operator, gives
\begin{align}
&S_{mn}(t,s,\tau) = G^{nn}_p(\tau) G^{mm}_p(-t) \det \left[ \left( 1-\hat{N} \right) \right. \nonumber \\
\label{SmnFormula}
&\left. +  e^{i h_d \tau}  e^{-i h_{d,n} \tau}  e^{i h_d s}  e^{i h_{d,m} t}  e^{-i h_0(t+s) } \hat{N}\right],
\end{align}
where $\hat{N} \equiv \left(1+ e^{H_0/k_B T } \right)^{-1}$ and $H_d = d^\dg_i \left( \hat{H}_d \right)_{ij} d_j$.  To handle spin, we let $m \rightarrow (m, \sigma)$ represent a combined site and spin index and replace the basis $\{i\}$ of Wannier orbitals with a spin-Wannier basis $\{ (i, \sigma) \}$.  The above determinant formula requires a quadractic Hamiltonian of the form $H = d^\dg_i h d_j$.  We map singlet pairing Hamiltonians with terms of the form $d^\dg_\uparrow d^\dg_\downarrow$ to the necessary form $d^\dg_\uparrow d_\downarrow$ via a particle-hole transformation $d^\dg_\downarrow \leftrightarrow d_\downarrow$.

The $4p$ Green functions in Eq.~(\ref{IndirectS}) appear to complicate the analysis of indirect RIXS but in fact simplify it by restricting the number of particle-hole excitations caused by the core hole.  Because the $4p$ band is highly dispersive the same-site Green function $G^{nn}_p(t)$ decays very rapidly -- it is unlikely that a $4p$ electron created at site $n$ will return except after very brief times.  Therefore the time intervals associated with $t$ and $\tau$ are effectively truncated much more than by the core hole lifetime alone.  This makes numerical integration less computationally expensive, but more importantly dramatically reduces the contribution of processes in which the core hole potential generates multiple particle-hole ``shake-up'' pairs.  Therefore (see below) indirect RIXS spectra can be interpreted in terms of single shake-up pairs and are not dominated by complicated processes involving multiple shake-ups.  A common source of confusion is the assumption that short intermediate state timescales $t$ and $\tau$ imply poor energy resolution.  However,  the times $t$ and $\tau$ are conjugate to the incident photon energy $\omega_i$, and indeed spectra are virtually featureless as a function of $\omega_i$.  However, the energy transfer $\Delta \omega$ is conjugate to the time $s$, during which there is no core hole and no $4p$ electron.  Hence resolution of $\Delta \omega$ is limited only by instrumental resolution.   This preserves the dispersion information of $\Delta \omega$ vs. $\Delta \vk$ that is fundamental to RIXS.  

\begin{figure*}
$
\begin{array}{cc}
\includegraphics[width=0.4\linewidth]{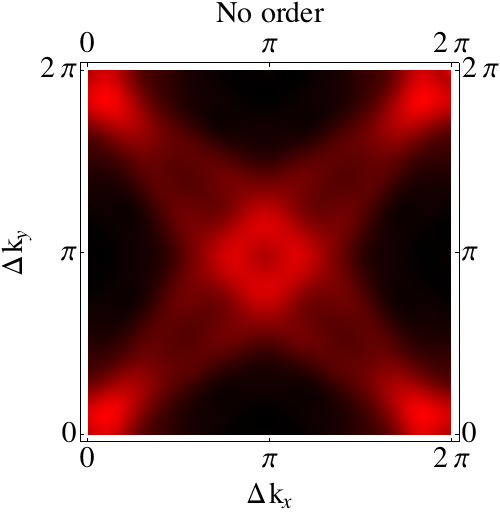} & \includegraphics[width=0.4\linewidth]{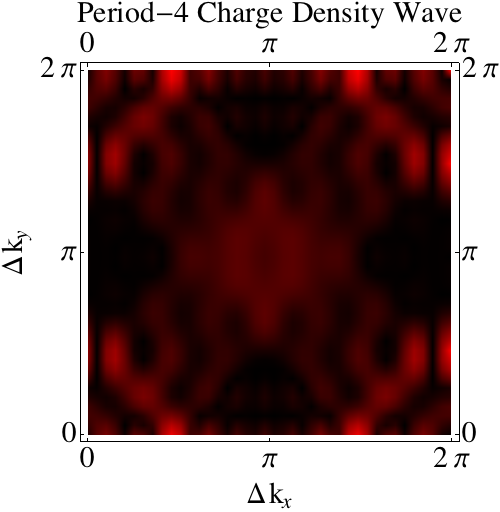}  \\
\includegraphics[width=0.4\linewidth]{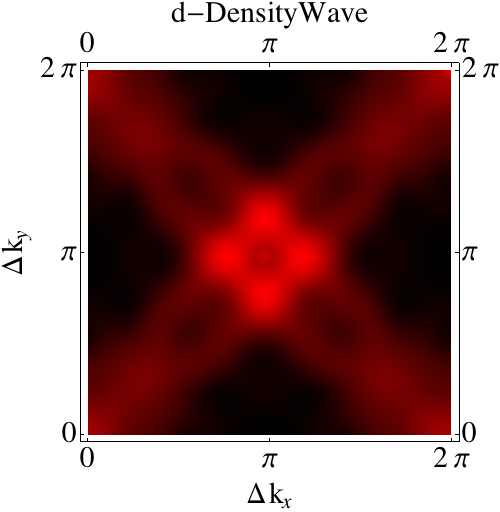} & \includegraphics[width=0.4\linewidth]{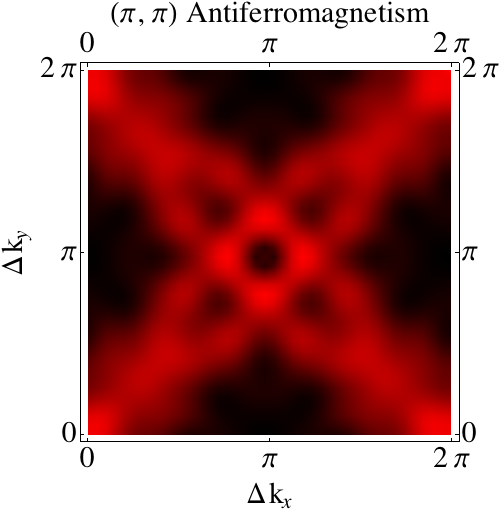}  
\end{array}
$
\caption{Indirect RIXS intensity (arbitrary units) versus momentum over entire Brillouin zone for underdoped Bi-2212 described by Hamiltonians $H_0$ (upper left) and mean-field Hamiltonians (clockwise from upper right): $H_0 + H_{CDW}$, $H_0 + H_{DDW}$, and $H_0 + H_{AF}$, with mean-field perturbations of amplitude $V_{CDW,DDW,AF} = 100$ meV.
Spectra are measured at energy transfer $\Delta \omega = 100$ meV and incident photon energy $\omega$ at the $4p$ threshold.  Here and elsewhere in this paper black denotes zero intensity and bright red denotes maximal intensity.}
\label{spectra}
\end{figure*}

\section{Main Results}
We are interested in whether indirect RIXS distinguishes different types of short-range order in hole-doped cuprates, particularly those that are hypothesized to exist in the pseudogap phase of underdoped cuprates above $T_c$.  In order to exploit the determinantal formalism, which requires a quadratic Hamiltonian, we will treat these orders as mean-field additions to the band structure Hamiltonian
\begin{equation}
H_0=\sum_{\vk, \sigma} \epsilon_\vk d^\dg_{\vk,\sigma} d_{\vk,\sigma},
\end{equation}
For concreteness we will use a single-band tight-binding dispersion $\epsilon_\vk = -2 t_1 (\cos (k_x) + \cos (k_y)) - 4 t_2 \cos (k_x) \cos (k_y) - 2 t_3 (\cos (2 k_x) + \cos (2 k_y) ) - 4 t_4 (\cos (2 k_x) \cos (k_y) + \cos (k_x) \cos (2 k_y) )$ with parameters fit to ARPES data: $(t_1, t_2, t_3, t_4) = (126, -36, 15, 1.5)$ meV for Bi-2212~\cite{Markiewicz2005}.  We assume an attractive contact potential $V_m = -U_c \sum_\sigma d^\dg_{m \sigma} d_{m \sigma}$ for the core hole, with $U_c = 5.0$ eV~\cite{Tsutsui1999, Tsutsui2003}.  To $H_0$ we add charge density wave (CDW), $d$-density wave, and antiferromagnetic (AF) orders:
\begin{align}
H_{CDW} =&  \sum_\vk V_{CDW} d^\dg_{\vk+\vQ} d_\vk \\
H_{DDW} =&  \sum_\vk V_{DDW}(\vk) d^\dg_{\vk+\vQ} d_\vk \\
\label{HSDW}
H_{AF} =&  \sum_\vk V_{AF} \left[ d^\dg_{\vk+\vQ,\up} d_{\vk,\up} - d^\dg_{\vk+\vQ,\dn} d_{\vk,\dn}  \right],
\end{align}
where the ordering wavevectors are $\vQ=(2 \pi/4,0)$ (CDW), $\vQ = (\pi,\pi)$ (DDW), and $\vQ = (\pi,\pi)$ (AF), and $V_{DDW}(\vk) = V_{DDW}(\cos k_x - \cos k_y)$.  We restrict our attention here to period-4 commensurate CDW order, but the qualitative features of the RIXS signal we present below are not specific to this wavevector.  The form of $H_{AF}$ in Eq. (\ref{HSDW}) is that of an alternating sublattice magnetization, which could occur in cuprates if residual local antiferromagnetic correlations persist after long-range antiferromagnetic order is destroyed by hole doping.  It is important to compare this form of AF to DDW because they have the same $(\pi,\pi)$ wavevector.  Thus we are able to study whether indirect RIXS is sensitive not only to the ordering wavevector but also to the form factor $V_{DDW}(\vk)$.  Another important distinction to note is that the DDW is orthogonal to conventional charge order -- while DDW is a form of translational symmetry breaking, the charge density due to DDW does not exhibit symmetry breaking.  We will therefore be able to reject any naive suspicion that indirect RIXS is only sensitive to order parameters that accompany a density distortion.

In direct RIXS experiments, and indeed in most spectroscopic experiments it is customary to present data in the form of lineshapes, that is, intensity versus $\Delta \omega$ and $\Delta \vk$ for momenta along some fixed cut in momentum space.  This makes sense for presenting the dispersion of collective modes.  However, in indirect RIXS of underdoped cuprates the fundamental excitations are a continuum of particle-hole pairs, the dispersion of which is not inherently interesting.  Rather, the indirect RIXS intensity measures the joint density of states of particles and holes, which are useful in that they reflect the overall fermiology over the entire Brillouin zone.  Thus, the most natural way to present indirect RIXS data is as plots of intensity versus momenta over the entire Brillouin zone for fixed $\Delta \omega$.

\begin{figure}
$
\begin{array}{c}
\includegraphics[width=0.8\linewidth]{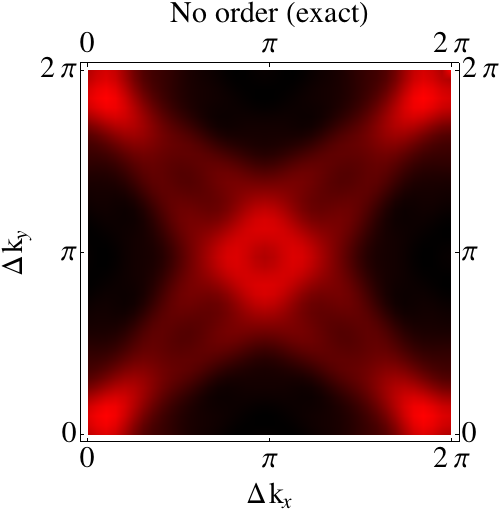} \\
\includegraphics[width=0.8\linewidth]{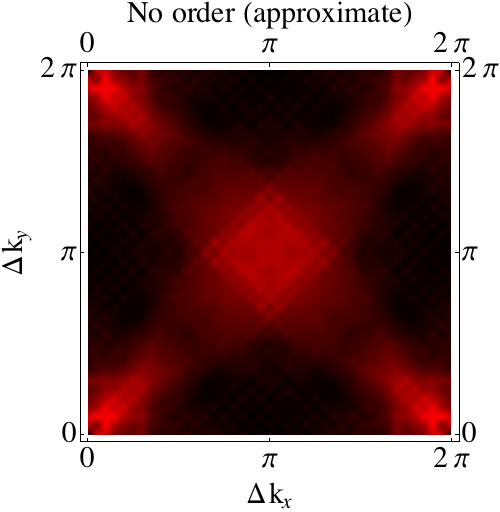} 
\end{array}
$
\label{compare}
\caption{Comparison of exact (Eqs.~(\ref{TimeDomain}) and~(\ref{SmnFormula})) and approximate (Eq.~(\ref{Approx}) formulas for momentum-dependent indirect RIXS intensity of unordered system for doping $p = 0.15$, energy transfer $\Delta \omega = 100$ meV.}
\end{figure}

In Fig. \ref{spectra} we examine the indirect RIXS spectrum in a state with no terms in $H_d$ other than the band structure $H_0$ and compare it to states in which various mean-field order parameters are added to $H_0$.  The intensity in the unordered state corresponds closely to the joint density of particle-hole pairs with total momentum $\Delta k$ and total energy $\Delta \omega$.  Besides the peaknear zero momentu transfer, the dominant feature is the peak at $\Delta k = (\pi,\pi)$ at energies several hundred meV and less.  This is due to the large density of states for both particles and holes near antinodal regions $(\pi,0)$ etc, which is caused by a saddle point in the dispersion.  The peak does not occur exactly at $(\pi,\pi)$ because the Fermi surface does not cross antinodal momenta $(0,\pi)$ and $(\pi,0)$, and thus there exists no antinodal-antinodal particle-hole pair with momentum $(\pi,0)-(0,\pi) = (\pi,\pi)$.  In a non-interacting system at half filling and only nearest-neighbor hopping there are van Hove singularities at antinodal momenta and we would expect the diamond-shaped Fermi surface to yield a RIXS maximum at $(\pi,\pi)$ for small energy transfers.  For the ordered systems we use order parameter amplitudes $V_{DDW}=100$ meV, $V_{CDW}=100$ meV, $V_{AF}=100$ meV, which are typical energy scales for the pseudogap of the cuprates' normal state.  These yield distinct changes in the RIXS spectra that allow experiments to distinguish them.  Of particular interest is that the $d$-density wave phase exhibits a clear signature distinct from other phases, including the antiferromagnetic phase which also has a wavevector of $(\pi,\pi)$.  As seen in Fig. \ref{spectra}, indirect RIXS at small $\Delta \omega$ of systems with DDW and AF order follows a similar pattern to unordered systems described by the band structure $H_0$: intensity maxima at zero momentum and around $(\pi,\pi)$, joined by arms running along the nodal directions $(k,k)$.  In the DDW system, the maximum near $(\pi,\pi)$ is strengthened relative to the maximum near $(0,0)$, while in the AF system intensity increases along the arms.  The behavior of the AF spectrum is directly related to the Fermi surface reconstruction due to ordering at a wavevector of $(\pi,\pi)$.  As long as $V_{AF}$ is not extremely strong, the bulk of the Fermi surface is replaced by large oval-shaped pockets that overlap the unreconstructed Fermi surface on one side and parallel it on the other.  From the point of view of the joint density of states of particle hole pairs, the effect is similar to a broadening of the Fermi surface.  In contrast, the $d$-density wave has the same $(\pi,\pi)$ wavevector, but the form factor $V_{DDW}(\vk)$ vanishes in the nodal direction and is maximal in the antinodal direction.  Hence the Fermi surface reconstruction due to $DDW$ order is restricted to the antinodal momenta near $(\pi,0)$ and $(0,\pi)$.  This explains why the ``arms'' of the indirect RIXS intensity pattern are not broadened as in the system with AF order.  The reason the intensity maximum at $(\pi,\pi)$ is strengthened is that the antinodal saddle point of the dispersion is buried inside the unreconstructed Fermi surface -- there are antinodal holes, but only \textit{near}-antinodal particles.  After reconstruction, there are more particles available near the saddle point.

\begin{figure*}
$
\begin{array}{ccc}
\includegraphics[width=0.3\linewidth]{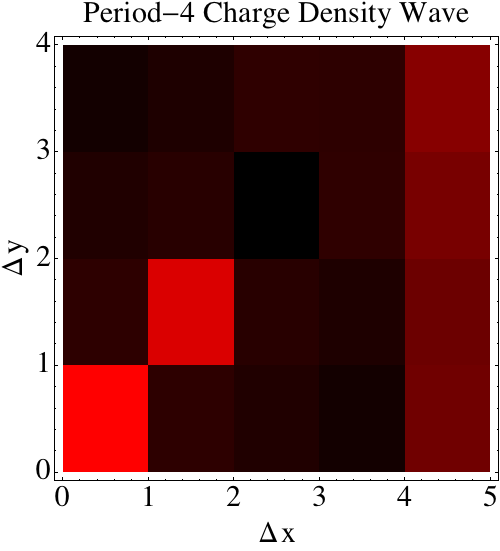} &
\includegraphics[width=0.3\linewidth]{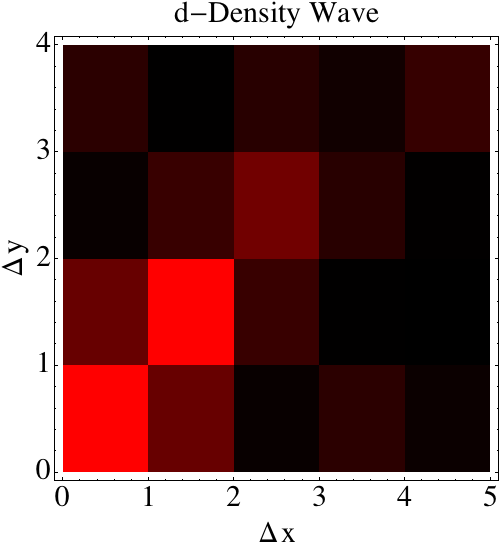} & 
\includegraphics[width=0.3\linewidth]{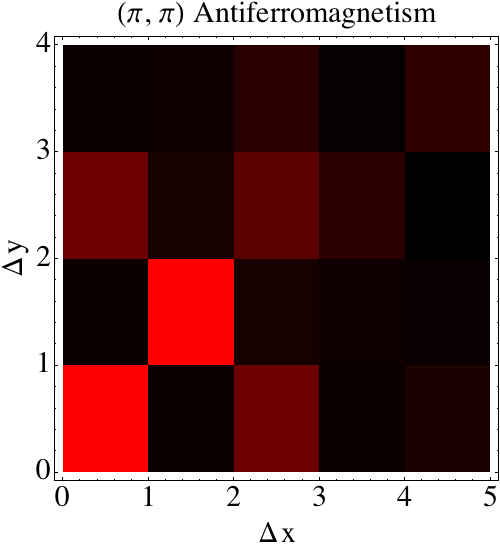}  
\end{array}
$
\caption{Fourier transforms of RIXS intensities from Fig.~\ref{spectra} exhibiting clear period-4 periodicity of the CDW system and checkerboard periodicity in the DDW and AF systems.}
\label{Fourier}
\end{figure*}

Unlike the previous examples, CDW order strongly modifies the intensity pattern of the orderless system.  One distinct feature is the appearance of maxima near $(0,2\pi/4)$, which is to be expected of Fermi surface reconstruction due to translational symmetry breaking with this wavevector.  This is a robust feature of RIXS at low energy transfers; indeed, at zero energy transfer this momentum becomes the elastic peak of the symmetry-broken system.  That is, the CDW by definition induces elastic scattering at wavevector $Q_{CDW}$, which generates low-energy particle hole pairs with total momentum $Q_{CDW}$.  Another obvious feature is the destruction of the intensity near $(\pi,\pi)$, which was a maximum for the unordered system as well as the DDW and AF systems.  We observe generally that a perturbation with wavevector $\vQ$ does not destroy the density of states of low energy particle-hole pairs with wavevector $\vQ$.  (Of course, the order yields a gap at energy scales of meV or tens of meV, which are much smaller than current energy scales measured by RIXS).  Thus DDW and AF order are in a sense ``compatible'' with the RIXS spectrum of the unordered system.  Ordering at a different wavevector, on the other hand, can and does drastically change the particle-hole joint density of states.  The clear qualitative difference between the CDW spectrum and those of DDW and AF systems, along with the maximum at $\vQ_{CDW}$, are telltale signs of translational symmetry breaking at a wavevector other than $(\pi,\pi)$.  However, one can discern different orders even more clearly with a complementary measurement:  Recall from above that the Keldysh-like two-site correlator $S_{mn}$ can be measured as the Fourier transform of intensity.  Specifically, the Fourier transform of $I(\Delta \vk,\omega,\Delta \omega)$ gives $S(\Delta \vr, \omega,\Delta \omega) \equiv \sum_{\vr_m - \vr_n = \Delta \vr} S_{mn}(\omega,\Delta \omega)$.  It is intuitively clear that $S(\Delta \vr)$ ought to have a spatial structure corresponding to that of $H$.  In Fig.~\ref{Fourier}, we see that the Fourier transform of RIXS intensity exhibits the spatial periodicity of $H$, that is, a checkerboard pattern for $DDW$ and $AF$ orders and a stripe at $\Delta x = 4$ for period-4 CDW.  (Even though our calculations were performed on $40 \times 40$ systems, we show only a few near-neighbor lattice sites in the Fourier-transformed spectra because $S_{mn}$ decays rapidly with separation between $\vr_m$ and $\vr_n$).

Thus, each of the orders we have considered have not only distinct patterns of indirect RIXS, but ``smoking gun'' signatures with simple interpretations.  There are, of course, many proposed pseudogap order parameters other than the CDW, DDW, and AF that we have considered here.  However, these three examples demonstrate that one can easily obtain robust, falsifiable predictions for the indirect RIXS spectrum of any candidate order.  Furthermore, these are among the most widely-proposed orders for the normal state of underdoped cuprates and it is encouraging that they exhibit such distinct patterns.

\section{Modelling RIXS with a Single Particle-Hole Pair}
As mentioned above, our exact results ought to be well-approximated by considering only a single shake-up pair in indirect RIXS.  It is straightforward to derive the RIXS intensity under this approximation for arbitrary bilinear Hamiltonians $H_d$ containing any combination of mean-field and impurity potentials.  We obtain
\begin{align}
I \propto& \sum_{\alpha,\beta} \left|
\tilde{V}_{\alpha,\beta}(\vk) \int \frac{g(\epsilon) d\epsilon}{(\omega-(\epsilon_\alpha-\epsilon_\beta -\epsilon + i \Gamma)(\omega - \epsilon + i \Gamma)} \right|^2 \nonumber \\
\label{Approx}
& \times n_f(\epsilon_\alpha)(1-n_f(\epsilon_\beta)) \delta(\epsilon_\alpha-\epsilon_\beta-\Delta \omega),  \\
\tilde{V}_{\alpha,\beta}(\vk) &= \sum_m e^{i \vk \cdot \vr_m} \braketop{\alpha}{V_m}{\beta} \nonumber
\end{align}
where $\ket{\alpha}$ and $\ket{\beta}$ are single-particle eigenstates of $H_d$, $g(\epsilon)$ is the 4$p$ density of states, and $\tilde{V}(\vk) = \sum_m e^{i \vk \cdot \vr_m} V_m$ is the Fourier transform of the core hole potential operator.  We show in Fig.~\ref{compare} that Eq.~(\ref{Approx}) gives results very similar to the exact formula Eq.~(\ref{SmnFormula}), which we used to generate the figures in this paper.  In addition to making explicit the connection between RIXS and the joint particle-hole density of states, Eq.~(\ref{Approx}) is also useful for analyzing the dependence of the RIXS signal on the incident photon energy $\omega$.  Here the resonance in $\omega$ is convoluted with the broad and featureless 4$p$ density of states.  The intensity is maximal when $\omega$ is in resonance to excite the core electron to the 4$p$ band minimum, where the group velocity vanishes and the $p$ electron is most likely to return to the core hole site.  Other than this feature the resonant factor yields little structure.  In particular, the overall shape of the spectrum as a function of $\Delta \vk$ and $\Delta \omega$ is unchanged as $\omega$ varies (to the point that $\omega = 10$ eV yields figures identical to those shown above, up to an overall scale), although the overall magnitude is strongly $\omega$-dependent.  In experiments one should tune $\omega$ to maximize the intensity but varying $\omega$ yields no useful information.  

\section{Alternative Analysis of RIXS Data}
\begin{figure*}
$
\begin{array}{cc}
\includegraphics[width=0.4\linewidth]{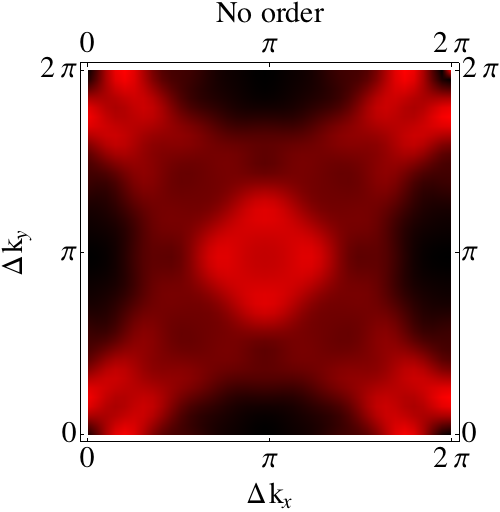} & \includegraphics[width=0.4\linewidth]{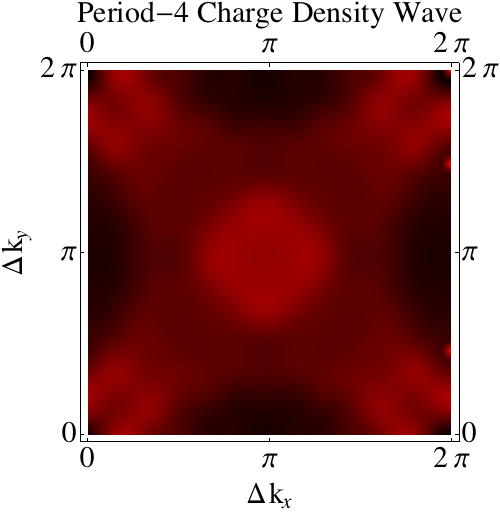}  \\
\includegraphics[width=0.4\linewidth]{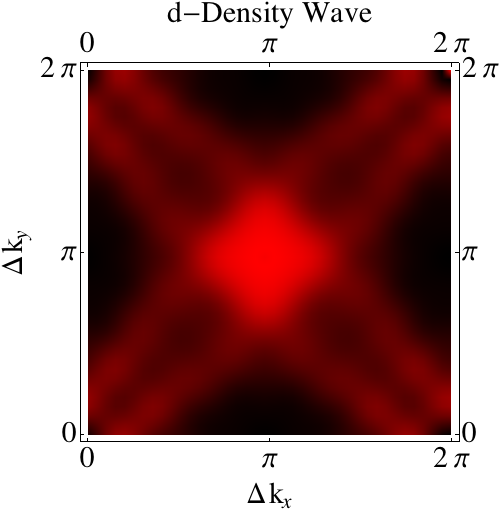} & \includegraphics[width=0.4\linewidth]{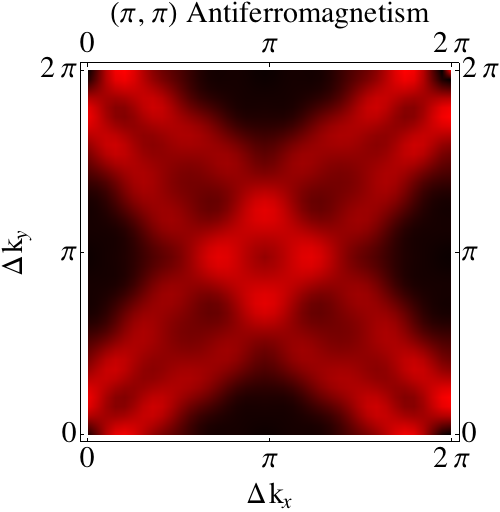}  
\end{array}
$
\caption{Same as Fig.~\ref{spectra} but with $\Delta \omega$ integrated from 100 meV to 500 meV.  Clockwise from top left: unordered state, CDW, AF, DDW.}
\label{integrated}
\end{figure*}
\begin{figure*}
$
\begin{array}{ccc}
\includegraphics[width=0.3\linewidth]{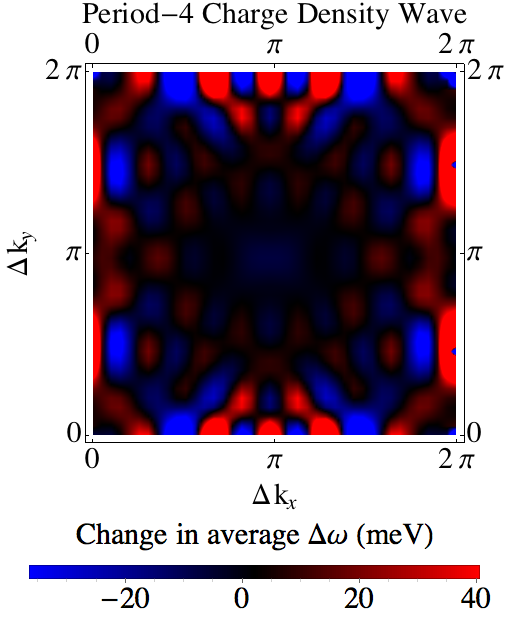} &
\includegraphics[width=0.3\linewidth]{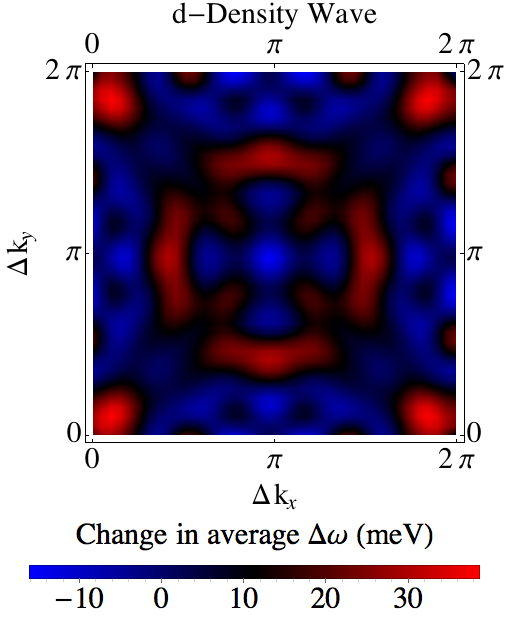} & 
\includegraphics[width=0.3\linewidth]{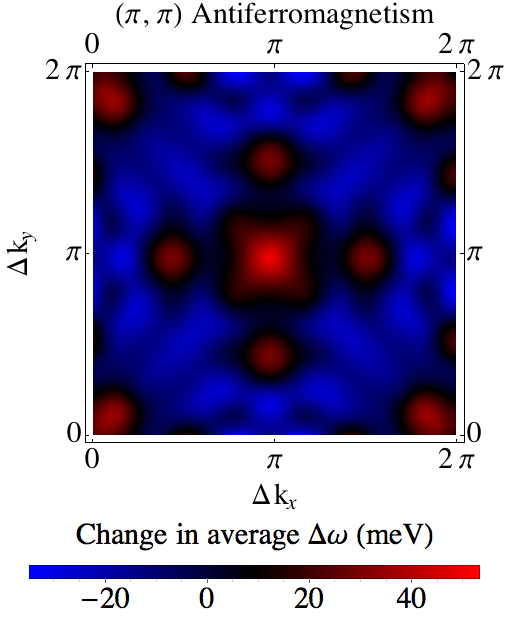}  
\end{array}
$
\caption{Change in first moments integrated over 100 meV $ \le \Delta \omega \le $ 500 meV relative to unordered first moments as function of momenta across the entire Brillouin zone.}
\label{average}
\end{figure*}
It is possible that in real samples, with the combined effects of disorder, inhomogeneity, and non-resonant scattering, it may be desirable to increase the statistical power and robustness of measurements.  One way to do this is to define appropriate averaged variables.  One natural choice is simply to average over a range of $\Delta \omega$:
\begin{equation}
\int_{\Delta \omega_1}^{\Delta \omega_2} I(\Delta \vk,\Delta \omega) d \Delta \omega,
\end{equation}
The most useful choice of the interval $[\Delta \omega_1,\Delta \omega_2]$ will be some range of small energy transfers with $\Delta \omega_1$ large enough to avoid the elastic peak but with $\Delta \omega_2$ small enough that the averaged quantity still reflects the redistribution of low-energy particles and holes due to Fermi surface reconstruction.  In Fig.~\ref{integrated} we plot this integrated intensity in the window 100 meV $ \le \Delta \omega \le$ 500 meV.  The same qualitative distinctions between spectra that appeared for fixed energy transfer $\Delta \omega = 100 $ meV are also present in the integrated spectra, namely, concentration of intensity near $\Delta \vk = (\pi,\pi)$ for DDW, increase of intensity along the ``arms'' $\Delta k = (k,k), \, 0 \le k \le \pi$ for AF order, and general diffuseness of intensity across the Brillouin zone for CDW.  Another measure is the first moment, that is, the averaged energy transfer $\Delta \omega$ weighted by the intensity:
\begin{equation}
{\rm 1st \, moment}(\Delta \vk)=\int \Delta \omega I(\Delta \vk,\Delta \omega) d \Delta \omega
\end{equation}
This measures the shift of particle-hole pairs to higher energies as gaps are opened at parts of the Fermi surface, and to lower energies as new parts of the Fermi surface arise.  In Fig.~\ref{average} we show the distinct patterns in the first moments for DDW, CDW, and AF orders relative to the unordered state.

{\it Summary}.--  We presented a formalism for treating exact band structures and core hole potentials in indirect RIXS.  We showed that indirect RIXS measures the joint density of states of particles and holes and is sensitive to perturbations to band structure due to the formation of local ordered states.

{\it Acknowledgements}.--  We gratefully acknowledge Dmitry Abanin and Daniel Podolsky for useful discussions and Peter Abbamonte for spurring our interest in resonant x-ray scattering.  We acknowlege support from Harvard-MIT CUA, AFOSR New Quantum Phases of Matter MURI, the ARO-MURI on Atomtronics, and the ARO MURI Quism program.

\bibliography{library}

\begin{thebibliography}{25}
\expandafter\ifx\csname natexlab\endcsname\relax\def\natexlab#1{#1}\fi
\expandafter\ifx\csname bibnamefont\endcsname\relax
  \def\bibnamefont#1{#1}\fi
\expandafter\ifx\csname bibfnamefont\endcsname\relax
  \def\bibfnamefont#1{#1}\fi
\expandafter\ifx\csname citenamefont\endcsname\relax
  \def\citenamefont#1{#1}\fi
\expandafter\ifx\csname url\endcsname\relax
  \def\url#1{\texttt{#1}}\fi
\expandafter\ifx\csname urlprefix\endcsname\relax\def\urlprefix{URL }\fi
\providecommand{\bibinfo}[2]{#2}
\providecommand{\eprint}[2][]{\url{#2}}

\bibitem[{\citenamefont{Timusk and Statt}(1999)}]{Timusk1999}
\bibinfo{author}{\bibfnamefont{T.}~\bibnamefont{Timusk}} \bibnamefont{and}
  \bibinfo{author}{\bibfnamefont{B.}~\bibnamefont{Statt}},
  \bibinfo{journal}{Reports on Progress in Physics}
  \textbf{\bibinfo{volume}{62}}, \bibinfo{pages}{61} (\bibinfo{year}{1999}),
  ISSN \bibinfo{issn}{0034-4885},
  \urlprefix\url{http://iopscience.iop.org/0034-4885/62/1/002}.

\bibitem[{\citenamefont{Kivelson et~al.}(1998)\citenamefont{Kivelson, Fradkin,
  and Emery}}]{Kivelson1998}
\bibinfo{author}{\bibfnamefont{S.~A.} \bibnamefont{Kivelson}},
  \bibinfo{author}{\bibfnamefont{E.}~\bibnamefont{Fradkin}}, \bibnamefont{and}
  \bibinfo{author}{\bibfnamefont{V.~J.} \bibnamefont{Emery}},
  \textbf{\bibinfo{volume}{393}}, \bibinfo{pages}{550} (\bibinfo{year}{1998}),
  ISSN \bibinfo{issn}{0028-0836},
  \urlprefix\url{http://dx.doi.org/10.1038/31177}.

\bibitem[{\citenamefont{Sachdev}(2000)}]{Sachdev2000}
\bibinfo{author}{\bibfnamefont{S.}~\bibnamefont{Sachdev}},
  \bibinfo{journal}{Science} \textbf{\bibinfo{volume}{288}},
  \bibinfo{pages}{475} (\bibinfo{year}{2000}), ISSN \bibinfo{issn}{00368075},
  \urlprefix\url{http://www.sciencemag.org/content/288/5465/475.abstract}.

\bibitem[{\citenamefont{Chakravarty et~al.}(2001)\citenamefont{Chakravarty,
  Laughlin, Morr, and Nayak}}]{Chakravarty2001}
\bibinfo{author}{\bibfnamefont{S.}~\bibnamefont{Chakravarty}},
  \bibinfo{author}{\bibfnamefont{R.}~\bibnamefont{Laughlin}},
  \bibinfo{author}{\bibfnamefont{D.}~\bibnamefont{Morr}}, \bibnamefont{and}
  \bibinfo{author}{\bibfnamefont{C.}~\bibnamefont{Nayak}},
  \bibinfo{journal}{Physical Review B} \textbf{\bibinfo{volume}{63}}
  (\bibinfo{year}{2001}), ISSN \bibinfo{issn}{0163-1829},
  \urlprefix\url{http://prb.aps.org/abstract/PRB/v63/i9/e094503}.

\bibitem[{\citenamefont{Kivelson et~al.}(2003)\citenamefont{Kivelson, Bindloss,
  Oganesyan, Tranquada, Kapitulnik, and Howald}}]{Kivelson2003a}
\bibinfo{author}{\bibfnamefont{S.~A.} \bibnamefont{Kivelson}},
  \bibinfo{author}{\bibfnamefont{I.~P.} \bibnamefont{Bindloss}},
  \bibinfo{author}{\bibfnamefont{V.}~\bibnamefont{Oganesyan}},
  \bibinfo{author}{\bibfnamefont{J.~M.} \bibnamefont{Tranquada}},
  \bibinfo{author}{\bibfnamefont{A.}~\bibnamefont{Kapitulnik}},
  \bibnamefont{and} \bibinfo{author}{\bibfnamefont{C.}~\bibnamefont{Howald}},
  \bibinfo{journal}{Reviews of Modern Physics} \textbf{\bibinfo{volume}{75}},
  \bibinfo{pages}{1201} (\bibinfo{year}{2003}), ISSN \bibinfo{issn}{0034-6861},
  \urlprefix\url{http://rmp.aps.org/abstract/RMP/v75/i4/p1201\_1}.

\bibitem[{\citenamefont{Norman et~al.}(2005)\citenamefont{Norman, Pines, and
  Kallin}}]{Norman2005}
\bibinfo{author}{\bibfnamefont{M.~R.} \bibnamefont{Norman}},
  \bibinfo{author}{\bibfnamefont{D.}~\bibnamefont{Pines}}, \bibnamefont{and}
  \bibinfo{author}{\bibfnamefont{C.}~\bibnamefont{Kallin}},
  \bibinfo{journal}{Advances in Physics} \textbf{\bibinfo{volume}{54}},
  \bibinfo{pages}{715} (\bibinfo{year}{2005}), ISSN \bibinfo{issn}{0001-8732},
  \urlprefix\url{http://www.tandfonline.com/doi/abs/10.1080/00018730500459906}.

\bibitem[{\citenamefont{Valla et~al.}(2006)\citenamefont{Valla, Fedorov, Lee,
  Davis, and Gu}}]{Valla2006a}
\bibinfo{author}{\bibfnamefont{T.}~\bibnamefont{Valla}},
  \bibinfo{author}{\bibfnamefont{A.~V.} \bibnamefont{Fedorov}},
  \bibinfo{author}{\bibfnamefont{J.}~\bibnamefont{Lee}},
  \bibinfo{author}{\bibfnamefont{J.~C.} \bibnamefont{Davis}}, \bibnamefont{and}
  \bibinfo{author}{\bibfnamefont{G.~D.} \bibnamefont{Gu}},
  \bibinfo{journal}{Science (New York, N.Y.)} \textbf{\bibinfo{volume}{314}},
  \bibinfo{pages}{1914} (\bibinfo{year}{2006}), ISSN \bibinfo{issn}{1095-9203},
  \urlprefix\url{http://www.sciencemag.org/content/314/5807/1914.abstract}.

\bibitem[{\citenamefont{Vojta}(2009)}]{Vojta2009b}
\bibinfo{author}{\bibfnamefont{M.}~\bibnamefont{Vojta}},
  \bibinfo{journal}{Advances in Physics} \textbf{\bibinfo{volume}{58}},
  \bibinfo{pages}{699} (\bibinfo{year}{2009}), ISSN \bibinfo{issn}{0001-8732},
  \urlprefix\url{http://www.informaworld.com/openurl?genre=article\&doi=10.1080/00018730903122242\&magic=crossref||D404A21C5BB053405B1A640AFFD44AE3}.

\bibitem[{\citenamefont{Berg et~al.}(2009)\citenamefont{Berg, Fradkin,
  Kivelson, and Tranquada}}]{Berg2009}
\bibinfo{author}{\bibfnamefont{E.}~\bibnamefont{Berg}},
  \bibinfo{author}{\bibfnamefont{E.}~\bibnamefont{Fradkin}},
  \bibinfo{author}{\bibfnamefont{S.~A.} \bibnamefont{Kivelson}},
  \bibnamefont{and} \bibinfo{author}{\bibfnamefont{J.~M.}
  \bibnamefont{Tranquada}}, \bibinfo{journal}{New Journal of Physics}
  \textbf{\bibinfo{volume}{11}}, \bibinfo{pages}{115004}
  (\bibinfo{year}{2009}), ISSN \bibinfo{issn}{1367-2630},
  \urlprefix\url{http://stacks.iop.org/1367-2630/11/i=11/a=115004?key=crossref.133e2c7edaa67659589460155b38f0bc}.

\bibitem[{\citenamefont{Baeriswyl et~al.}(2009)\citenamefont{Baeriswyl,
  Eichenberger, and Menteshashvili}}]{Baeriswyl2009}
\bibinfo{author}{\bibfnamefont{D.}~\bibnamefont{Baeriswyl}},
  \bibinfo{author}{\bibfnamefont{D.}~\bibnamefont{Eichenberger}},
  \bibnamefont{and}
  \bibinfo{author}{\bibfnamefont{M.}~\bibnamefont{Menteshashvili}},
  \bibinfo{journal}{New Journal of Physics} \textbf{\bibinfo{volume}{11}},
  \bibinfo{pages}{075010} (\bibinfo{year}{2009}), ISSN
  \bibinfo{issn}{1367-2630},
  \urlprefix\url{http://stacks.iop.org/1367-2630/11/i=7/a=075010}.

\bibitem[{\citenamefont{Davis and Lee}(2013)}]{Davis2013}
\bibinfo{author}{\bibfnamefont{J.~C.~S.} \bibnamefont{Davis}} \bibnamefont{and}
  \bibinfo{author}{\bibfnamefont{D.-H.} \bibnamefont{Lee}},
  \bibinfo{journal}{Proceedings of the National Academy of Sciences of the
  United States of America} \textbf{\bibinfo{volume}{110}},
  \bibinfo{pages}{17623} (\bibinfo{year}{2013}), ISSN
  \bibinfo{issn}{1091-6490},
  \urlprefix\url{http://www.pnas.org/content/110/44/17623}.

\bibitem[{\citenamefont{Sachdev}(2003)}]{Sachdev2003}
\bibinfo{author}{\bibfnamefont{S.}~\bibnamefont{Sachdev}},
  \bibinfo{journal}{Reviews of Modern Physics} \textbf{\bibinfo{volume}{75}},
  \bibinfo{pages}{913} (\bibinfo{year}{2003}), ISSN \bibinfo{issn}{0034-6861},
  \urlprefix\url{http://rmp.aps.org/abstract/RMP/v75/i3/p913\_1}.

\bibitem[{\citenamefont{Kotani and Shin}(2001)}]{Kotani2001}
\bibinfo{author}{\bibfnamefont{A.}~\bibnamefont{Kotani}} \bibnamefont{and}
  \bibinfo{author}{\bibfnamefont{S.}~\bibnamefont{Shin}},
  \bibinfo{journal}{Reviews of Modern Physics} \textbf{\bibinfo{volume}{73}},
  \bibinfo{pages}{203} (\bibinfo{year}{2001}), ISSN \bibinfo{issn}{0034-6861},
  \urlprefix\url{http://link.aps.org/doi/10.1103/RevModPhys.73.203}.

\bibitem[{\citenamefont{Abbamonte et~al.}(2012)\citenamefont{Abbamonte, Demler,
  {S\'{e}amus Davis}, and Campuzano}}]{Abbamonte2012}
\bibinfo{author}{\bibfnamefont{P.}~\bibnamefont{Abbamonte}},
  \bibinfo{author}{\bibfnamefont{E.}~\bibnamefont{Demler}},
  \bibinfo{author}{\bibfnamefont{J.}~\bibnamefont{{S\'{e}amus Davis}}},
  \bibnamefont{and} \bibinfo{author}{\bibfnamefont{J.-C.}
  \bibnamefont{Campuzano}}, \bibinfo{journal}{Physica C: Superconductivity}
  \textbf{\bibinfo{volume}{481}}, \bibinfo{pages}{15} (\bibinfo{year}{2012}),
  ISSN \bibinfo{issn}{09214534},
  \urlprefix\url{http://dx.doi.org/10.1016/j.physc.2012.04.006}.

\bibitem[{\citenamefont{Nozi\`{e}res and Abrahams}(1974)}]{Nozieres1974}
\bibinfo{author}{\bibfnamefont{P.}~\bibnamefont{Nozi\`{e}res}}
  \bibnamefont{and} \bibinfo{author}{\bibfnamefont{E.}~\bibnamefont{Abrahams}},
  \bibinfo{journal}{Physical Review B} \textbf{\bibinfo{volume}{10}},
  \bibinfo{pages}{3099} (\bibinfo{year}{1974}), ISSN \bibinfo{issn}{0556-2805},
  \urlprefix\url{http://prb.aps.org/abstract/PRB/v10/i8/p3099\_1}.

\bibitem[{\citenamefont{Benjamin
  et~al.}(2013{\natexlab{a}})\citenamefont{Benjamin, Klich, and
  Demler}}]{Benjamin2013a}
\bibinfo{author}{\bibfnamefont{D.}~\bibnamefont{Benjamin}},
  \bibinfo{author}{\bibfnamefont{I.}~\bibnamefont{Klich}}, \bibnamefont{and}
  \bibinfo{author}{\bibfnamefont{E.}~\bibnamefont{Demler}},
  p.~\bibinfo{pages}{6} (\bibinfo{year}{2013}{\natexlab{a}}),
  \eprint{1312.6642}, \urlprefix\url{http://arxiv.org/abs/1312.6642}.

\bibitem[{\citenamefont{Tsutsui et~al.}(1999)\citenamefont{Tsutsui, Tohyama,
  and Maekawa}}]{Tsutsui1999}
\bibinfo{author}{\bibfnamefont{K.}~\bibnamefont{Tsutsui}},
  \bibinfo{author}{\bibfnamefont{T.}~\bibnamefont{Tohyama}}, \bibnamefont{and}
  \bibinfo{author}{\bibfnamefont{S.}~\bibnamefont{Maekawa}},
  \bibinfo{journal}{Physical Review Letters} \textbf{\bibinfo{volume}{83}},
  \bibinfo{pages}{3705} (\bibinfo{year}{1999}), ISSN \bibinfo{issn}{0031-9007},
  \urlprefix\url{http://prl.aps.org/abstract/PRL/v83/i18/p3705\_1}.

\bibitem[{\citenamefont{Benjamin et~al.}(2014)\citenamefont{Benjamin, Klich,
  and Demler}}]{Benjamin2014}
\bibinfo{author}{\bibfnamefont{D.}~\bibnamefont{Benjamin}},
  \bibinfo{author}{\bibfnamefont{I.}~\bibnamefont{Klich}}, \bibnamefont{and}
  \bibinfo{author}{\bibfnamefont{E.}~\bibnamefont{Demler}},
  \bibinfo{journal}{Physical Review Letters} \textbf{\bibinfo{volume}{112}},
  \bibinfo{pages}{247002} (\bibinfo{year}{2014}), ISSN
  \bibinfo{issn}{0031-9007},
  \urlprefix\url{http://link.aps.org/doi/10.1103/PhysRevLett.112.247002}.

\bibitem[{\citenamefont{Scalapino and Sugar}(1981)}]{Scalapino1981}
\bibinfo{author}{\bibfnamefont{D.}~\bibnamefont{Scalapino}} \bibnamefont{and}
  \bibinfo{author}{\bibfnamefont{R.}~\bibnamefont{Sugar}},
  \bibinfo{journal}{Physical Review Letters} \textbf{\bibinfo{volume}{46}},
  \bibinfo{pages}{519} (\bibinfo{year}{1981}), ISSN \bibinfo{issn}{0031-9007},
  \urlprefix\url{http://link.aps.org/doi/10.1103/PhysRevLett.46.519}.

\bibitem[{\citenamefont{Scalettar et~al.}(1986)\citenamefont{Scalettar,
  Scalapino, and Sugar}}]{Scalettar1986}
\bibinfo{author}{\bibfnamefont{R.}~\bibnamefont{Scalettar}},
  \bibinfo{author}{\bibfnamefont{D.}~\bibnamefont{Scalapino}},
  \bibnamefont{and} \bibinfo{author}{\bibfnamefont{R.}~\bibnamefont{Sugar}},
  \bibinfo{journal}{Physical Review B} \textbf{\bibinfo{volume}{34}},
  \bibinfo{pages}{7911} (\bibinfo{year}{1986}), ISSN \bibinfo{issn}{0163-1829},
  \urlprefix\url{http://link.aps.org/doi/10.1103/PhysRevB.34.7911}.

\bibitem[{\citenamefont{Klich}(2003)}]{Klich2004}
\bibinfo{author}{\bibfnamefont{I.}~\bibnamefont{Klich}}, in
  \emph{\bibinfo{booktitle}{Quantum Noise in Mesoscopic Physics}}, edited by
  \bibinfo{editor}{\bibfnamefont{Y.}~\bibnamefont{Nazarov}}
  (\bibinfo{publisher}{Springer}, \bibinfo{year}{2003}),
  \urlprefix\url{http://www.springer.com/physics/quantum+physics/book/978-1-4020-1239-6}.

\bibitem[{\citenamefont{Abanin and Levitov}(2005)}]{Abanin2005}
\bibinfo{author}{\bibfnamefont{D.}~\bibnamefont{Abanin}} \bibnamefont{and}
  \bibinfo{author}{\bibfnamefont{L.}~\bibnamefont{Levitov}},
  \bibinfo{journal}{Physical Review Letters} \textbf{\bibinfo{volume}{94}},
  \bibinfo{pages}{186803} (\bibinfo{year}{2005}), ISSN
  \bibinfo{issn}{0031-9007},
  \urlprefix\url{http://prl.aps.org/abstract/PRL/v94/i18/e186803}.

\bibitem[{\citenamefont{Benjamin
  et~al.}(2013{\natexlab{b}})\citenamefont{Benjamin, Abanin, Abbamonte, and
  Demler}}]{Benjamin2013}
\bibinfo{author}{\bibfnamefont{D.}~\bibnamefont{Benjamin}},
  \bibinfo{author}{\bibfnamefont{D.}~\bibnamefont{Abanin}},
  \bibinfo{author}{\bibfnamefont{P.}~\bibnamefont{Abbamonte}},
  \bibnamefont{and} \bibinfo{author}{\bibfnamefont{E.}~\bibnamefont{Demler}},
  \bibinfo{journal}{Physical Review Letters} \textbf{\bibinfo{volume}{110}},
  \bibinfo{pages}{137002} (\bibinfo{year}{2013}{\natexlab{b}}), ISSN
  \bibinfo{issn}{0031-9007},
  \urlprefix\url{http://link.aps.org/doi/10.1103/PhysRevLett.110.137002}.

\bibitem[{\citenamefont{Markiewicz et~al.}(2005)\citenamefont{Markiewicz,
  Sahrakorpi, Lindroos, Lin, and Bansil}}]{Markiewicz2005}
\bibinfo{author}{\bibfnamefont{R.~S.} \bibnamefont{Markiewicz}},
  \bibinfo{author}{\bibfnamefont{S.}~\bibnamefont{Sahrakorpi}},
  \bibinfo{author}{\bibfnamefont{M.}~\bibnamefont{Lindroos}},
  \bibinfo{author}{\bibfnamefont{H.}~\bibnamefont{Lin}}, \bibnamefont{and}
  \bibinfo{author}{\bibfnamefont{A.}~\bibnamefont{Bansil}},
  \bibinfo{journal}{Physical Review B} \textbf{\bibinfo{volume}{72}},
  \bibinfo{pages}{054519} (\bibinfo{year}{2005}), ISSN
  \bibinfo{issn}{1098-0121},
  \urlprefix\url{http://prb.aps.org/abstract/PRB/v72/i5/e054519}.

\bibitem[{\citenamefont{Tsutsui et~al.}(2003)\citenamefont{Tsutsui, Tohyama,
  and Maekawa}}]{Tsutsui2003}
\bibinfo{author}{\bibfnamefont{K.}~\bibnamefont{Tsutsui}},
  \bibinfo{author}{\bibfnamefont{T.}~\bibnamefont{Tohyama}}, \bibnamefont{and}
  \bibinfo{author}{\bibfnamefont{S.}~\bibnamefont{Maekawa}},
  \bibinfo{journal}{Physical Review Letters} \textbf{\bibinfo{volume}{91}}
  (\bibinfo{year}{2003}), ISSN \bibinfo{issn}{0031-9007},
  \urlprefix\url{http://prl.aps.org/abstract/PRL/v91/i11/e117001}.

\end{thebibliography}

\end{document}